\documentclass[twocolumn,preprintnumbers,showpacs, amsmath,amssymb,superscriptaddress,prl]{revtex4}

\usepackage{graphicx}
\usepackage{dcolumn}
\usepackage{bm}
\usepackage{amsmath}
\usepackage{amsfonts}
\usepackage{hyperref}
\usepackage{amssymb}
\begin{document}

\title{Transport of Orbital-Angular-Momentum Entanglement through a Turbulent Atmosphere}

\author{Bart-Jan Pors}
\affiliation{Huygens Laboratory, Leiden University, P.O.\ Box
9504, 2300 RA Leiden, The Netherlands}
\author{C. H. Monken} \affiliation{Huygens Laboratory, Leiden
University, P.O.\ Box 9504, 2300 RA Leiden, The Netherlands}
\affiliation{Departamento de F\'{i}sica, Universidade Federal de
Minas Gerais, Caixa Postal 702, Belo Horizonte, MG 30123-970,
Brazil}
\author{Eric R. Eliel} \affiliation{Huygens Laboratory, Leiden University,
P.O.\ Box 9504, 2300 RA Leiden, The Netherlands}
\author{J.P. Woerdman} \affiliation{Huygens Laboratory, Leiden University,
P.O.\ Box 9504, 2300 RA Leiden, The Netherlands}

\begin{abstract}

We demonstrate experimentally how orbital-angular-momentum entanglement of two photons evolves under influence of atmospheric turbulence. We find that the quantum channel capacity is surprisingly robust: Its typical horizontal decay distance is of the order of 2 kilometers, demonstrating the potential of photonic orbital angular momentum for free-space quantum communication in a metropolitan environment.

\end{abstract}


\maketitle

Quantum communication by means of entangled photon pairs allows for an intrinsically secure transmission of data, by distributing the pairs via a free-space or fiber channel to distant parties \cite{Ekert}. Most popular is polarization entanglement, which has dimensionality 2. Higher dimensionalities can be achieved using orbital-angular-momentum (OAM) entanglement \cite{Mair, Kawase, PorsPRL2008} or energy-time entanglement \cite{Riedmatten, Ali-Khan}; this route provides for a larger channel capacity and an increased security against eavesdroppers \cite{Durt, Brukner2}. However, the performance of a practical high-dimensional quantum channel is an open issue. Here, we address this issue for the case of OAM entanglement distribution via a free-space channel.

For quantum communication to be of practical relevance, it is imperative that the entanglement between the photons carrying the information survives over a reasonably long propagation distance. Entanglement distribution over fiber-based transmission lines has proven to maintain coherence over tens of kilometers \cite{Poppe, Zhang2, Salart}. However, the use of free-space links cannot be obviated when considering such purposes as airplane and satellite quantum links or hand-held communication devices \cite{Resch, Peng, Ursin}.

The increased quantum-channel capacity that is available when encoding the information in the OAM of the entangled photons was anticipated to be severely limited in a practical free-space link, due to atmospheric turbulence that causes wave front distortions.  Several theoretical studies have addressed this aspect \cite{Paterson_OAMturbulence, Gibson, Gopaul, Gbur, Anguita, Walborn2, Tyler}, but there is no unanimity on exactly \textit{how} sensitive OAM entanglement is to atmospheric perturbations. So far, no experimental verdict has been obtained to clarify this issue.

In this Letter, we present the first such experiment. We start with bipartite OAM entanglement of dimensionality 6, and demonstrate how the corresponding quantum correlations evolve when one of the photons traverses a turbulent atmosphere, emulated by controlled mixing of cold and hot air. Our experimental results are in excellent agreement with our theoretical model, which is based on a Kolmogorov description of atmospheric turbulence. Specifically, we show how increasing strengths of turbulence degrade the Shannon dimensionality, which was introduced in Ref. \cite{PorsPRL2008} as a simple measure to quantify the channel capacity. Scaling up our system to real-life dimensions, we find that its typical horizontal decay distance is about 2 km.

\begin{figure}[!b]
\includegraphics[angle=0,width=7.5truecm]{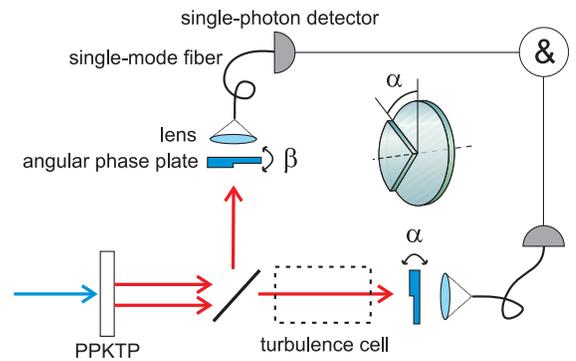}
\caption{\label{fig:1} Experimental setup. A type-I PPKTP crystal emits two frequency-degenerate photons ($\lambda = 826$ nm) that are entangled in their OAM degree of freedom. A beam splitter serves to separate the twin photons spatially. The entanglement is analyzed by two angular-phase-plate projectors, variably oriented at $\alpha$ and $\beta$, respectively, which are linked to a coincidence circuit. Each phase plate has one elevated quadrant sector with optical thickness $\lambda/2$ (inset). In one of the beam lines we place a turbulence cell. }
\end{figure}
Our experimental setup is depicted in Fig. \ref{fig:1}. The PPKTP crystal emits correlated photon pairs with complementary OAM, in a state of the form $|\Psi\rangle = \sum_{l, p} c_{p}^l \left|l, p\right\rangle \left|-l, p\right\rangle$ \cite{Torres, Footnote2}. Here, $|l, p \rangle$ indicates the Schmidt mode containing one photon with orbital angular momentum $l\hbar$, with $p$ the radial mode index, and we can write $\langle \mathbf{r}|l, p \rangle = \langle r|l, p \rangle e^{i l \theta}/\sqrt{2\pi}$. The total number of entangled azimuthal modes, the so-called angular Schmidt number $K$, is of order 30 \cite{Exter_ModCount}. The correlated photons are spatially separated by a 50/50 beam splitter.

The entanglement is analyzed by means of two state projectors, which are composed of an angular phase plate that is lens-coupled to a single-mode fiber and a single-photon counter. The two phase plates are identical and carry a purely azimuthal variation of their optical thickness: they have one elevated quadrant sector with an optical thickness that is $\lambda/2$ larger than that of the remainder of the plate (see inset Fig. \ref{fig:1}). The two phase plates can be rotated around their normals over an angle $\alpha$ and $\beta$, respectively. The detection state $|A(\alpha)\rangle$ (or $|B(\beta)\rangle$) of one such analyzer is a high-dimensional superposition of OAM modes, the relative phases of which depend on the orientation of the plate. It can be written as $\langle\mathbf{r}|A(\alpha)\rangle = (2/w_0) \exp(-r^2/w_0^2) \sum_l \sqrt{\lambda_l} e^{i l (\theta + \alpha)}$, where the Gaussian factor describes the fiber mode profile with field radius $w_0$, and the summation over the orbital-angular-momentum states describes the phase imprint imparted by the phase plate. When rotating the quadrant phase plate over $2\pi$ rad, the analyzer scans a mode space of dimensionality $D = 6$ \cite{Pors}.

\begin{figure}[]
\includegraphics[angle=0,width=8truecm]{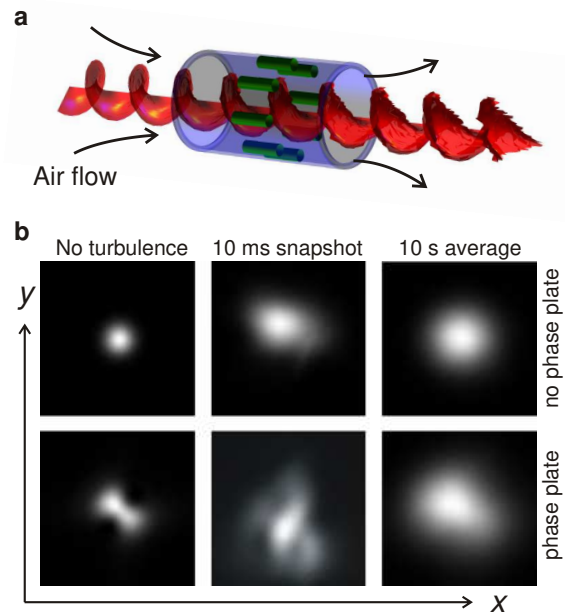}
\caption{\label{fig:2} Beam corruption after passage through the turbulence cell. (a) Impression of how an OAM eigenmode, having a helical wavefront, gets distorted when transiting the turbulence cell. The cell consists of a 7 cm long, 26 mm diameter glass tube, containing several resistors that produce up to 60 W of heat. A gentle flow of room temperature air is driven through the tube. (b) Far-field intensity patterns of the analyzer, which is fed backwards with diode laser light at 826 nm. The analyzer is equipped with no phase plate (top row), or quadrant-sector phase plate (with its sector aligned along the Cartesian axes) (bottom row). The diffraction limited patterns (left column) get perturbed when turbulence is switched on (middle column): for mild turbulence, the dominant effect is a randomly evolving beam deflection; for the more severe turbulence conditions used here ($w_0/r_0 = 0.65$), the beam profile can get significantly distorted. Taking a 10 s time average reveals an isotropic beam broadening (right column). The apparent asymmetry along the diagonal in the bottom left and right windows is due to the 3\% discrepancy of the quadrant phase step from the ideal value of $\pi$. }
\end{figure}

In one of the beam lines, we place a turbulence cell where cold and hot air are mixed to bring about random variations of the refractive index that vary over time (Fig. \ref{fig:2}(a)). We can tune the strength of the turbulence by varying the heating power and air flow through the cell. Similar cells have been used as a realistic emulation of atmospheric turbulence \cite{Keskin}. Figure \ref{fig:2}(b) gives an impression of the cell's functioning: We inject one of the analyzers backwards with diode laser light and monitor the beam, which traverses the turbulence cell, in the far field. We do this for two cases; the analyzer is equipped with no phase plate (top row), or with the quadrant-sector phase plate (bottom row). We observe that the input beams (left column) become deformed by the refractive index fluctuations, as can be seen when taking a 10 ms snapshot (middle column). Time averaging these fluctuations over 10 s reveals a beam broadening that is spatially isotropic (right column).

We describe our cell by the Kolmogorov theory of turbulence \cite{Tatarski}. This standard model treats the optical effects of the atmosphere at any moment as a random phase operation $e^{i \phi(\mathbf{r})}$, the time evolution of which follows a Gaussian distribution. It is conveniently described in terms of its coherence function, given by
\begin{equation}
    \label{Eq.1}
    \left\langle  e^{i \phi(\mathbf{r}_1)- i \phi(\mathbf{r}_2)}\right\rangle_t =
    e^{-\frac{1}{2}6.88\left[ \frac{\mathbf{r}_1 - \mathbf{r}_2}{r_0} \right] ^{5/3}},
\end{equation}
where $\langle \ldots \rangle_t$ denotes averaging over time \cite{Fried}. The relevant parameter in this model is the Fried parameter $r_0$, being the transverse distance over which the beam profile gets distorted by approximately 1 rad of root-mean-square phase aberration \cite{Fried}. In the absence of turbulence $r_0 \rightarrow \infty$, but when turbulence becomes stronger, the spatial coherence is reduced and hence $r_0$ shortens. From the Gaussian beam broadening in Fig. \ref{fig:2}(b) (top row) we can determine the relation between the Fried parameter and the $1/e$ beam size $w_0$,
\begin{equation}
    \label{Eq.2}
    \frac{w_0}{r_0} = \frac{\sqrt{(w_{le}/w_{dl})^2-1}}{3.0},
\end{equation}
with $w_{dl}$ and $w_{le}$ the $1/e$ far-field radius of the diffraction limited beam and long-exposure broadened beam, respectively \cite{Fante}.

\begin{figure}[]
\includegraphics[angle=0,width=8.3truecm]{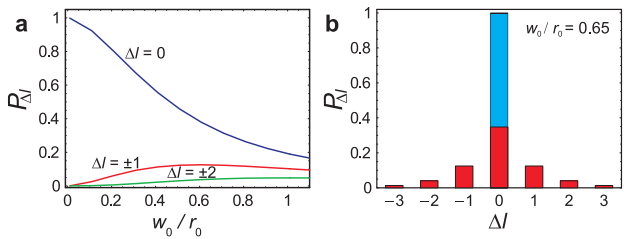}
\caption{\label{fig:3} Mode scattering due to turbulence. (a) Survival probability of an analyzer's OAM eigenmode $l = l_0$ as a function of turbulence strength (blue). The red and green curves denote turbulence-induced coupling probabilities to neighboring modes for $\Delta l = \pm 1$ and $\Delta l = \pm2$, respectively. (b) Spreading of the $l_0$ OAM eigenmode (blue bar) over its neighbors for $w_0/r_0 = 0.65$ (red bars).}
\end{figure}

We calculated the effect of Kolmogorov turbulence on our analyzer state $|A(\alpha)\rangle$. The blue curve in Fig. \ref{fig:3}(a) shows the survival probability of an OAM eigenmode $l = l_0$ upon passing through a turbulent atmosphere as described by Eq. (\ref{Eq.1}). The survival probability degrades gradually for increasing turbulence strength. We note that this decay depends on the ratio $w_0/r_0$ only and not on the specific OAM eigenvalue $l_0$, provided that the propagation distance $L$ is small compared to the diffraction length $z_R = \pi w_0^2/\lambda$. Furthermore, the turbulence produces a coupling between the orthogonal OAM modes, leading to a non-vanishing mode overlap between the $l_0$ eigenmode and its neighbors $\Delta l = \pm1$ (red) and $\Delta l = \pm2$ (green). A different perspective on this mode mixing is presented in histogram Fig. \ref{fig:3}(b), which shows how an OAM eigenmode (blue bar) spreads out over its neighboring azimuthal modes for $w_0/r_0 = 0.65$ (red bars). We note that normalization is not preserved, because some intensity is scattered to radial modes that are not sustained by the single-mode fiber. This illustrates the importance of taking into account the radial content of the generated two-photon state and the analyzers' detection states when dealing with OAM modes in the presence of turbulence.

\begin{figure}[]
\includegraphics[angle=0,width=8truecm]{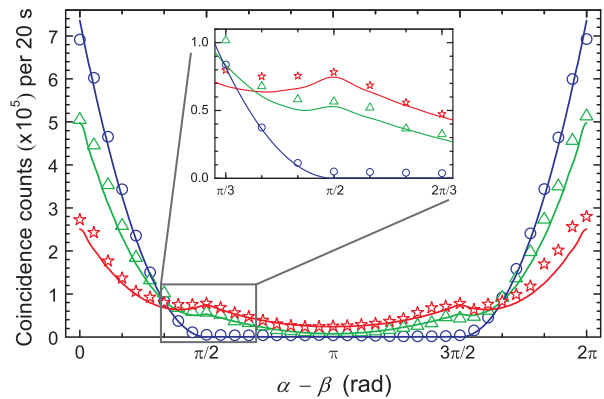}
\caption{\label{fig:4} Survival of OAM entanglement under influence of turbulence. Experimental coincidence rates (data points) and theoretical predictions (curves) obtained with two quadrant-sector phase plates for: no turbulence (blue), $w_0/r_0 = 0.30$ (green) and $w_0/r_0 = 0.65$ (red). The inset shows a blow-up of the wiggles around $\alpha-\beta = \pi/2$.}
\end{figure}

In the experiment, the phase plates are rotated around their normals, and the photon coincidence probability
\begin{equation}
    \label{Eq.3}
    P(\alpha-\beta) = \left\langle |\langle A(\alpha)|\langle B(\beta)|
    \hat{S}_A |\Psi\rangle|^2 \right\rangle_t
\end{equation}
is recorded as a function of their independent orientations. Here, the time-averaged behaviour of the turbulence scattering operator $\hat{S}_A$ is known by means of its coherence function Eq. (\ref{Eq.1}): $\left\langle \hat{S}_A|A(\alpha)\rangle \langle A(\alpha)| \hat{S}_A^\dagger \right\rangle_t = \int d\mathbf{r}_1 d\mathbf{r}_2 |\mathbf{r}_1 \rangle \langle \mathbf{r}_1|A(\alpha)\rangle \langle A(\alpha )|\mathbf{r}_2\rangle \langle \mathbf{r}_2| \left\langle e^{i \phi(\mathbf{r}_1)- i \phi(\mathbf{r}_2)}\right\rangle_t$. Figure \ref{fig:4} shows our main experimental results. In the absence of turbulence, we observe a piecewise-parabolic coincidence curve (blue circles), i.e. the coincidence rate follows a parabolic dependence for $|\alpha - \beta| \leq \pi/2$ and is zero elsewhere \cite{PorsPRL2008}. The coincidence rate depends on the relative orientation of the phase plates only. We have investigated how the coincidence rates evolve for 6 turbulence strengths, two of them shown in Fig. \ref{fig:4}: $w_0/r_0 = 0.30$ (green triangles) and $w_0/r_0 = 0.65$ (red stars). The latter strength was also used for Fig. \ref{fig:2}(b) and \ref{fig:3}(b). Note that the 20 s integration time assures isotropic sampling of the wavefront fluctuations (see Fig. \ref{fig:2}(b)). We observe a partial ``smoothening" of the coincidence curve, which is excellently described by our theoretical predictions based on Eqs. (\ref{Eq.1}) and (\ref{Eq.3}), without any fit parameter. The turbulence-induced wiggles at $|\alpha - \beta| = \pi/2$ are reproduced remarkably well (see inset).

Figure \ref{fig:4} shows that the coherence of the entangled state is partly conserved. We quantify its robustness in terms of the Shannon dimensionality $D$, which is a measure of the quantum channel capacity \cite{PorsPRL2008}. It is an operationally defined measure and gives the effective number of modes the combined analyzers have access to when scanning over their possible settings, \textit{viz.} the phase-plate orientations. For two identical analyzers in the absence of turbulence, we can express $D$ in terms of their \textit{pure} detection state operators $\rho_A$ and $\rho_B$, where $\rho_A = |A(\alpha)\rangle \langle A(\alpha)|$, as $D = 1/\mathrm{Tr} (\langle \rho_A \rangle_\alpha \langle \rho_B \rangle_\beta) = 1/\mathrm{Tr} (\langle \rho_A \rangle_\alpha)^2$. Here, $\langle \rho_A \rangle_\alpha$ is the density matrix averaged over all phase plate orientations $\alpha$.

In the presence of turbulent scattering, however, the detection state becomes randomly time dependent: $\rho_A = \rho_A(t)$. The relevant detection state operator is therefore not $\rho_A$, but rather $\langle \rho_A \rangle_t$, i.e., the density matrix averaged over time. In general, $\langle\rho_A\rangle_{t}$ is no longer a single-mode projector, but a mixed positive operator. In other words, when averaging over the random fluctuations, the detection state becomes multimode, which degrades the analyzer's modal resolution. Therefore, in the presence of turbulence, the Shannon dimensionality for \textit{mixed} detection states is given by
\begin{equation}
    \label{Eq.4}
    D = \frac{\rm{Tr}\left(\langle\rho\rangle_t\right)^2}{\rm{Tr}\left(\langle\rho\rangle_{t, \alpha}\right)^2},
\end{equation}
where $\langle\rho\rangle_{t, \alpha}$ denotes the average of the detection state operator $\rho$ over time $t$ and orientation $\alpha$. The inverse of the numerator of Eq. (\ref{Eq.4}) can be interpreted as the effective number of modes captured by the analyzer for fixed orientation $\alpha$. Alternatively, it also represents the purity $P = \mathrm{Tr} (\langle \rho \rangle_t)^2 \leq 1$ of the detection state, whose evolution follows the survival probability discussed in Fig. \ref{fig:3}(a) and is independent of the specific phase plates in use. Moreover, the numerator gives the maximum coincidence rates obtained at the peaks in Fig. \ref{fig:4}, up to a constant scaling factor \cite{Footnote3}.

\begin{figure}[]
\includegraphics[angle=0,width=8truecm]{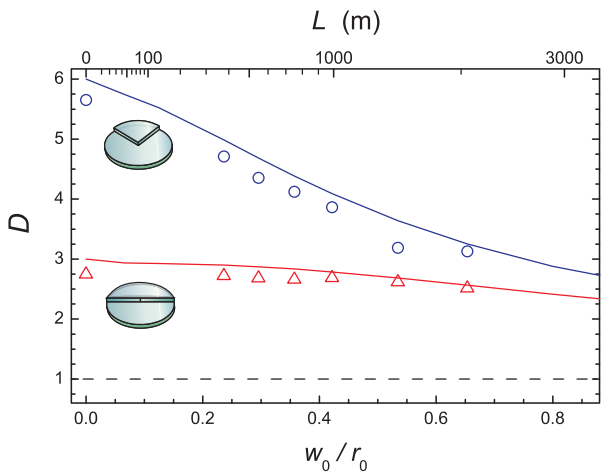}
\caption{\label{fig:5} Decay of Shannon dimensionality. Experimental (data points) and theoretical (curves) dimensionality as a function of turbulence strength, for two quadrant-sector phase plates (circles) and two half-sector phase plates (triangles). The turbulence strength is expressed in the ratio $w_0/r_0$ (lower abscissa) and in the corresponding real-life propagation distance $L$ (upper abscissa).}
\end{figure}

Experimentally, $D$ can straightforwardly be extracted from the coincidence curves in Fig. \ref{fig:4}: $D = 2\pi N_{max}/A$, where $N_{max}$ is the maximum coincidence rate and $A$ is the area underneath the curve. Figure \ref{fig:5} shows how $D$ evolves for increasing turbulence strength according to theory and experiment. In the absence of turbulence, we find an experimental value $D = 5.7$ vs. a theoretical prediction $D = 6$ \cite{PorsPRL2008}. As the turbulence strength increases, the modal resolution of the analyzers degrades, constraining the dimensionality to smaller values, ultimately to $D = 1$. The number of communication modes is reduced by $\sim 50\%$ to $D = 3.1$ when $w_0/r_0 = 0.65$ (theory $D = 3.3$). Considering the severity of the wave front distortions (see Fig. \ref{fig:2}(b)), we conclude that the channel capacity is surprisingly robust. For comparison, we also plotted our experimental results obtained with two half-sector phase plates, having one semicircle phase shifted by $\pi$ (see inset in Fig. \ref{fig:5}). For this case we observe that the dimensionality, initially at a value $D = 3$, decays considerably more slowly. This indicates that the resilience to atmospheric turbulence is very sensitive to the nature of the OAM superposition state, an aspect also noted in Ref. \cite{Anguita}. We expect that the search for such optimal states will lead to even more tenacity of the channel capacity.

The results presented here allow us to estimate the horizontal propagation distance $L$ that can be reached for free-space quantum communication outside the laboratory, since the Kolmogorov theory (Eq. (\ref{Eq.1})) used to describe our data is also a fair description of a real-life atmosphere \cite{Tatarski}. Let us consider the 50\% decay level of the channel capacity, roughly reached in our experiment for $w_0/r_0 = 0.65$ ($D = 3.1$). For horizontal propagation, the Fried parameter can be expressed as $r_0 = 3.02(k^2 L C_n^2)^{-3/5}$ \cite{Fried2}. Assuming moderate ground-level perturbations ($C_n^2 = 10^{-14}$ m$^{-2/3}$) \cite{Johnston}, a wavelength $\lambda$ = 1550 nm in the transmission window of the atmosphere and a beam size $w_0$ = 6 cm, we find a propagation length of 2 km (satisfying the requirement $L<z_R$). This is sufficient for quantum key distribution in a metropolitan environment, offering the advantages of high-dimensional entanglement as compared to the case of 2D polarization entanglement. We note that this distance could be enhanced significantly if one were to incorporate additional adaptive optics techniques \cite{Levine}.

We acknowledge valuable discussions with Steven Habraken, Laurent Jolissaint and Remco Stuik. CHM acknowledges financial support from the Brazilian agencies CNPq and CAPES. This project received funding from the Stichting voor Fundamenteel Onderzoek der Materie (FOM) and from the EU Seventh Framework Programme HIDEAS (grant agreement no. 221906).


\begin{thebibliography}{33}
\bibitem[{Foo({\natexlab{a}})}]{Footnote2}
\bibinfo{note}{Within the mode space our analysers have access to, we can
  safely approximate $c_{p}^l$ to be independent of $l$.}
\bibitem[{Foo({\natexlab{b}})}]{Footnote3}
\bibinfo{note}{We note that in our experiment we have turbulence in one arm
  only. For this case, the Shannon dimensionality can be generalized as $D =
  \rm{Tr}(\langle \rho_A \rangle_t \rho_B)/\rm{Tr}(\langle \rho_A
  \rangle_{t,\alpha} \langle \rho_B \rangle_{\beta})$. It can be shown that
  this reduces to Eq. (4) when one has similar but weaker turbulence in both
  arms.}
\end{thebibliography}

\end{document}